\documentclass[aps,prb,showpacs,twocolumn,groupedaddress,amssymb]{revtex4}

\usepackage{bm}
\usepackage{graphics}
\begin{document}

\title{Inhomogeneity-induced second-order phase transitions in Potts model on hierarchical lattices.}

\author{P. N. Timonin}
\email{timonin@aaanet.ru}
\affiliation{Physics Research Institute at Rostov State University
344090, Rostov - on - Don, Russia}

\date{\today}

\begin{abstract}
The thermodynamics of the $q$-state Potts model with arbitrary $q$ on a class of hierarchical lattices is considered. Contrary to the case of the crystal lattices, it has always the second-order phase transitions. The analytical expressions fo the critical indexes are obtained, their dependencies on the structural lattice pararmeters are studied and the scailing relations among them are establised. The structural criterion of the inhomogeneity-induced transformation of the transition order is suggested. The application of the results to a description of critical phenomena in the dilute crystals and substances confined in porous media is discussed.
\end{abstract}


\maketitle

The investigations of phase transitions on the hierarchical lattices have begun with the development of the Migdal-Kadanoff renormalization group method \cite{1,2} in which such lattices emerged as approximants of the ordinary crystal ones \cite{3}. Later the variety of hierarchical lattices with noninteger dimensionalities were introduced \cite{4, 5} thus being the models of fractal structures. As fractals are ubiquitous in disordered media such as dilute crystals and porous materials \cite{6}, the studies of  phase transitions in them are of great interest. While the structures of hierarchical lattices are not random they have broad distributions of coordination numbers and characteristic lengths being in this respect the useful models of random systems. Indeed the studies of the second-order transition in the Ising model on hierarchical lattices have shown the dependence of the critical indexes on the structural characteristics \cite{7,8} in close analogy with that found in experiments on disordered crystals under the variations of disorder strength \cite{9}.

The influence of structural inhomogeneity on the first-order transitions in hierarchical fractals also has qualitative similarity to that found in the numerical studies of such transitions in the Potts models in dilute crystals \cite{10,11,12,13}  and porous media \cite{14,15}. Thus the effect of  transformation of the first - order transitions  into the second-order ones found in Refs. \onlinecite{10,11,12,13,14,15} exists also in the $q$-state Potts model ($q = 4,10$) on two types of hierarchical lattices with fractal dimensions $d > 2 $ \cite{16}. This unusual phenomenon is hard to explain in the framework of the standard phenomenology, describing the smearing of first-order transitions in random media (diminishing or complete vanishing of the jumps of thermodynamic parameters) as a result of appearance of inhomogeneous two-phase state near transition \cite{17}. Such physical picture can not explain the nature of the instability which occurs in the inhomogeneous systems and results in the divergence of correlation length and critical susceptibility \cite{10,11,12,13,14,15}.

Meanwhile the experimental studies of transitions in liquid crystals \cite{18, 19} and antiferromagnet $MnO$ \cite{20}, confined into porous media, corroborate the possibility of the transition order transformation under the influence of inhomogeneities. The transformation of the structural first-order transition $O_h \to D_{4h}$ into the symmetry forbidden second-order transition takes place in the magnetite $Fe_3O_4$ under $Zn$ doping \cite{21}. Also the change of the transition order is observed in mixed crystals $(KBr)_{1-x}(KCN)_x$ \cite{22, 23} at the ferroelastic transition from the cubic phase to the orthorhombic one. Such transition in the ideal crystals is always of the first-order type \cite{24}, but in $(KBr)_{1-x}(KCN)_x$ it becomes of the second-order type at $x = 0.65, 0.7$ \cite{22} and $x = 0.73$ \cite{23} with elastic module $C_{44}$ going to zero at the transition point \cite{23}.

Thus the experiments and numerical studies of realistic models show that inhomogineity not only can smooth the first-order jumps of thermodynamic parameters but can also induce their second-order singularities. The elucidation of the nature of this phenomenon and development of its theoretical description could be very important for the understanding the mechanisms of phase transitions in random media as well as for the numerous practical applications of disordered materials. The task posed before theory is to determine the classes of first-order transitions and types of random media in which the second-order singularities appear and those in which usual jump smearing take place owing to the existence of intermediate inhomogeneous phase \cite{17,25}.  Another task is to determine the critical indexes for inhomogeneity-induced second-order transitions and to study their dependence on the structural characteristics of random media.

In solving these problems essential help can be obtained from the studies of spin models on such simplified imitations of real inhomogeneous systems as hierarchical lattices, the thermodynamics of which allows exact analytical description in some cases\cite{16}. Indeed the results of Ref. \onlinecite{16} along with the investigations of Potts model on random scale-free graphs \cite{26, 27} are the only analytical evidences of the possibility of transformation of first-order transitions into second-order ones in inhomogeneous systems.

Here we may note that the choice of Potts model for these studies in Refs. \onlinecite{10,11,12,13,14,15,16,26,27} as the simplest model having first-order transitions on translationally-invariant lattices with $d=2,3$ is also stipulated by the existence of its numerous physical realizations. Among them there are structural transitions in monolayers ($q = 3, 4$), transitions in the cubic ferromagnets in external field and in liquid mixtures ($q = 3$), see review Ref. \onlinecite{28} and references therein. There are also a number of ferroelastic transitions, described by the Potts model, e.g. transition $O_h \to D_4h$ in spinels of $Ni Cr_2 O_4$ and $Fe_3O_4$ type and superconductors $Nb_3 Sn$ and $V_3 Si$ ($q=3$) or charge ordering transitions in $Yb_4 As_3$ \cite{29} and alloys of $Mg_3 Cd$ type ($q=4$), see Table IV.4 in Ref. \onlinecite{24}.

Here we consider transitions in $q$ - state Potts model with arbitrary $q$ on the two-parametric family of hierarchical lattices with fractal dimensions $d >1$ and average coordination number $\bar z$ between $2$ and $4$. Using analytical approach differing from that of Ref. \onlinecite{16}, it is possible to show that second order transitions take place at all $q, d$ and $\bar z$, to obtain analytical expressions for critical indexes, to study their dependence on lattice structural parameters and to establish the scaling relations among them. The results obtained allow to suggest the structural criterion for the appearance of inhomogeneity-induced second-order transitions. Finally we discuss the application of present results to the description of critical anomalies at phase transitions in dilute crystals and substances confined in porous media.

\section{Geometrical characteristics of the hierarchical lattices}
\begin{figure}
\includegraphics{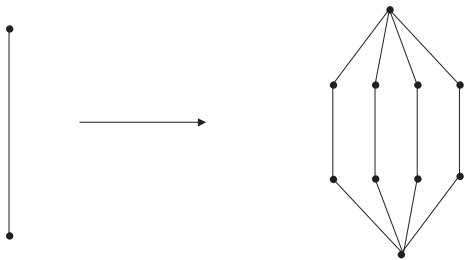}
\caption{\label{Fig.1}Graphical representation of the hierarchical lattice construction, $m=3$ and $n=4$.}
\end{figure}

The procedure for construction of the hierarchical lattices we will consider here is depicted in Fig.\ref{Fig.1}. It consists in sequential substitutions of every bond by the $n \ge 2$ chains having $m \ge 2$ bonds. 

At every step the number of bonds in the lattice becomes $B=mn$ times greater, so at $k$-th step we have $B^k$ bonds in it. The number of sites $N_k$ one can find from the recurrence relation
\[
N_k  = BN_{k - 1}- B - n + 2
\]
from which we have under the initial condition $N_0 = 2$
\begin{equation}
N_k  = \frac{{B - n}}{{B - 1}}B^k  + \frac{{n - 1}}{{B - 1}} + 1.
\label{eq:1}
\end{equation}
Introducing the average coordination number for infinite lattice, $\bar z \equiv \mathop {\lim }\limits_{k \to \infty } \frac{{2B^k }}{{N_k }}$ we get from Eq. (\ref{eq:1})
\begin{equation}
\bar z = 2\frac{B - 1}{B - n}.
\label{eq:2}
\end{equation}
Thus $\bar z$ can vary from 2 (at $m \to \infty $) up to 4 (at $m = 2,{\rm  }n \to \infty $). At $k$-th step the largest distence between lattice cites is $m^k$ so $N_k  \sim \left(m^k \right)^d $ at $k \to \infty $, where
\[
d = \frac{\ln B}{\ln m}
\]
is the fractal dimension of the lattice. Apparently, $1 < d < \infty $. 

Considering $\bar z$ and $d$ as independent parameters from inequalities $m \ge 2$, $n \ge 2$ we get
\begin{equation}
2 < \bar z < 4\left( {1 - 2^{ - d} } \right).
\label{eq:3}
\end{equation}

Let us consider the distributions of cite coordination numbers in such lattices. The coordination numbers acquire the values $z_k  = 2n^k $ and the number of sites having them in the lattice of the level $l$ is
\[
s_k  = \left( {B - n} \right)B^{l - k - 1} ,{\rm   0} \le k \le l - 1
\]
Also two basic sites have the coordination numbers $z_l  = n^l $, so $s_l = 2$. One can easily verify that  $\sum\limits_{k = 0}^l {s_k }  = N_l $ and $\sum\limits_{k = 0}^l {s_k z_k } = 2B_l $. Hence we have the following distribution function for coordination numbers in the infinite lattice
\[
W\left( z \right) = \mathop {\lim }\limits_{l \to \infty } \sum\limits_{k = 0}^l {\frac{{s_k }}{{N_l }}\delta \left( {z - z_k } \right)} 
\]
\[
 = \frac{{B - 1}}{B}\left( {\frac{2}{z}} \right)^{\frac{d}{{d - 1}}} \sum\limits_{k = 0}^\infty  {\delta \left( {z - 2n^k } \right)} 
\]
Thus $W\left( z \right)$ has power law dependence with exponent greater than $1$. Analogous distributions but with more dense sequence of $z_k$ ($z_k=k$) are used for modeling the scale-free random graphs \cite{26,27}. Potts model with $q\ge 1$ on such graphs has second - order transition with mean-field singularities if exponent in $W\left( z \right)$ is less than $3$ \cite{26}, which is related to the divergence of moment $\left\langle {z^2 } \right\rangle $. In our case all moments $\left\langle {z^r } \right\rangle $ with $r \ge d/\left( {d - 1} \right)$ diverge, while at lower $r$
\[
\left\langle {z^r } \right\rangle  = 2^r \frac{{B - 1}}{{B - n^r }}.
\]
Yet we show below that on the hierarchical lattices considered the transition is always of the second-order type and the properties of $W\left( z \right)$ influence only the values of the critical indexes.

\section{Recurrence relations, thermodynamic potential and correlations in Potts model}
The partition function of the $q$-state Potts model on the hierarchical lattices considered can be obtain introducing for each bond the factor
\begin{equation}
Z_0 \left( {\sigma ,\sigma '} \right) = \exp \left[ {K\delta _{\sigma ,\sigma '}  + \frac{h}{2}\left( {\delta _{\sigma ,1}  + \delta _{\sigma ',1} } \right)} \right]. 
\label{eq:4}
\end{equation}
Here $\sigma, \sigma'$ are the Potts spins on the connected cites, $K = J/T$,  $h$ - external field. Summing the obtained expression over the values of Potts spins $\sigma  = \left\{ {1,2,...,q} \right\}$ we get the full partition function.

Introducing the partial partition functions, $Z_l \left( {\sigma, \sigma'} \right)$, for the level $l$ lattice summed over all spins except $\sigma ,\sigma '$ on the basic sites, one can get the recurrence relations for them\cite{1,2,3}
\begin{equation}
Z_{l + 1} \left( {\sigma,\sigma'} \right) = \left[ {\left( \hat Z_l \right)_{\sigma,\sigma '}^m } \right]^n \exp \frac{h}{2}\left( {1 - n} \right)\left( \delta _{\sigma ,1}  + \delta _{\sigma ',1} \right),
\label{eq:5}
\end{equation}

Thus the next level partial partition function is obtained as $m$-th power of the matrix $Z_l \left( {\sigma ,\sigma '} \right)$ with subsequent exponentiation of each its element to the $n$-th power. The exponential factor in Eq. (\ref{eq:5}) deletes extra powers of $\exp h\left( \delta _{\sigma ,1}  + \delta _{\sigma ',1} \right)/2$.  Using Eqs. (\ref{eq:4}, \ref{eq:5}) one can find full partition function in the thermodynamic limit $l \to \infty $. 

From Eqs. (\ref{eq:4}, \ref{eq:5}) it follows that $Z_l \left( \sigma, \sigma'\right)$ can be represented as	
\begin{eqnarray}
 Z_l\left(\sigma,\sigma'\right) = a_{1l}\delta_{\sigma,1}\delta_{\sigma',1}+a_{2l}\left(1-\delta_{\sigma,1}\right)\left(1-\delta_{\sigma',1}\right)/(q - 1) 
\nonumber
\\
 + c_l\left[\delta_{\sigma,\sigma'}-\delta_{\sigma,1}\delta_{\sigma',1}-
\left(1-\delta_{\sigma,1}\right)\left(1-\delta_{\sigma',1}\right)/(q - 1)\right]
\label{eq:6}  
\\
 + b_l\left(\delta_{\sigma ,1}+\delta_{\sigma ',1}+\delta_{\sigma,1} \delta_{\sigma',1} \right).  
\nonumber 
\end{eqnarray}
From Eq. \ref{eq:6} it follows that the matrix $Z_l \left( {\sigma ,\sigma '} \right)$ has two nondegenerate eigenvalues,
\begin{equation}
\lambda _{ \pm l}  = \frac{1}{2}\left( {a_{1,l}  + a_{2,l} } \right) \pm \sqrt {\frac{1}{4}\left( {a_{1,l}  - a_{2,l} } \right)^2  + \left( {q - 1} \right)b_l^2 } 
\label{eq:7}
\end{equation}
and the eigenvalue $c_l $ with the degeneracy $q-2$. The transformation of the coefficients in the Eq. \ref{eq:6} when the matrix $Z_l \left( {\sigma ,\sigma '} \right)$ is exponentiated to the $m$-th power can be represented using Eq. \ref{eq:7} in the form
\begin{eqnarray}
a'_{1l}  = \frac{1}{2}\left( {\lambda _{ + l}^m  + \lambda _{ - l}^m } \right) + \frac{{\zeta _l }}{2}\left( {a_{1l}  - a_{2l} } \right),\nonumber\\
a'_{2l}  = \frac{1}{2}\left( {\lambda _{ + l}^m  + \lambda _{ - l}^m } \right) - \frac{{\zeta _l }}{2}\left( {a_{1l}  - a_{2l} } \right),\label{eq:8}\\
b'_l  = \zeta _l b_l,
\qquad
c'_l  = c_l^m.\nonumber
\end{eqnarray}
\[
\zeta _l  \equiv \left(\lambda _{ + l}^m  - \lambda _{ - l}^m \right)/\left(\lambda _{ + l}  - \lambda _{ - l} \right).
\]
Then the recurrence relations for coefficients corresponding to Eq. (\ref{eq:5}) are
\begin{eqnarray}
a_{1,l + 1}  = e^{ - h\left( {n - 1} \right)} \left( {a'_{1l} } \right)^n, 
\qquad
b_{l + 1}  = e^{ - h\left( {n - 1} \right)/2} \left( {b'_l } \right)^n, \nonumber \\
a_{2,l + 1}  = \left( {\frac{{a'_{2l}  + \left( {q - 2} \right)c'_l }}{{q - 1}}} \right)^n  + \left( {q - 2} \right)\left( {\frac{{a'_{2l}  - c'_l }}{{q - 1}}} \right)^n \label{eq:9} \\
c_{l + 1}  = \left( {\frac{{a'_{2l}  + \left( {q - 2} \right)c'_l }}{{q - 1}}} \right)^n  - \left( {\frac{{a'_{2l}  - c'_l }}{{q - 1}}} \right)^n \nonumber
\end{eqnarray}

According to Eq. (\ref{eq:4}) the initial conditions for these relations are
\begin{eqnarray}
a_{1,0}  = e^{K + h},
\qquad
a_{2,0}  = e^K  + q - 2,\nonumber\\
b_0  = e^{h/2},
\qquad
c_0  = e^K  - 1 \label{eq:10}
\end{eqnarray}

Solving Eqs. (\ref{eq:8}, \ref{eq:9}) one can find Potts partition functions with different boundary conditions on the basic lattice sites. Thus we get for free boundary conditions adding the missing fields $h/2$ to basic sites
\begin{eqnarray}
Z_l^{\left( f \right)}  = \sum\limits_{\sigma ,\sigma '} {e^{h\left( {\sigma  + \sigma '} \right)/2} Z_l \left( {\sigma ,\sigma '} \right)}\nonumber\\
= e^h a_{1l}  + \left( {q - 1} \right)\left( {a_{2l}  + 2e^{h/2} b_l } \right)
\label{eq:11}
\end{eqnarray}
For the periodic boundary conditions identifying the basic sites we have
\begin{equation}
Z_l^{\left( p \right)}  = \sum\limits_\sigma  {Z_l \left( {\sigma ,\sigma } \right)}  = a_{1l}  + a_{2l}  + \left( {q - 2} \right)c_l .
\label{eq:12}
\end{equation}
At last for the boundary conditions with fixed spins $\sigma  = 1$ on the basic sites 
\begin{equation}
Z_l^{\left( 1 \right)}  = Z_l \left( {1,1} \right) = a_{1l}. 
\label{eq:13}
\end{equation}

In the absence of long-range order all these partition functions give in thermodynamic limit the same values of the thermodynamic potential per site. At $h \ne 0$ their determination is rather hard task, yet the potential and its field derivatives can be more easily obtained at $h=0$ by the analytical methods. This is sufficient for determination of the transition order and description of critical anomalies.

In this paragraph we obtain the expression for the thermodynamic potential in zero field. In this case we have from Eqs. (\ref{eq:8}, \ref{eq:9})	
\[
a_{1l}  = b_l  + c_l 
\qquad
a_{2l}  = a_{1l}  + \left( {q - 2} \right)c_l ,
\]
so in zero field there are only two independent coefficients. Introducing
\[
K_l  \equiv \ln \frac{{a_{1l} }}{{b_l }},
\]
we get the well-known relation \cite{30}
\begin{eqnarray}
e^{K_{l + 1} }  = f\left( {e^{K_l } } \right),
\\
f\left( x \right) \equiv \left[ {1 + \frac{{q\left( {x - 1} \right)^m }}{{\left( {x - 1 + q} \right)^m  - \left( {x - 1} \right)^m }}} \right]^n \nonumber
\label{eq:14}
\end{eqnarray}
The second recurrence relation at $h = 0$ is:
\begin{equation}
b_{l + 1}  = g_{l + 1} b_l^B ,
\qquad
g_l  \equiv \frac{{\left( {e^{K_{l - 1} }  - 1} \right)^B }}{{\left( {e^{K_l /n}  - 1} \right)^n }}
\label{eq:15}
\end{equation}
From Eq. (\ref{eq:15}) and the definition of $K_l $ we get the expression for the thermodynamic potential in zero field:
\begin{equation}
F =  - T\mathop {\lim }\limits_{l \to \infty } N_l^{ - 1} \ln Z_l^{\left( f \right)}  =  - \frac{{\bar z}}{2}T\sum\limits_{k = 1}^\infty  {B^{ - k} \ln g_k }.
\label{eq:16}
\end{equation}
To derive Eq. (\ref{eq:16}) we take into account that $K_l B^{ - l}  \to 0$ when $l \to \infty $. 

The relations in Eq. (\ref{eq:14}) have one stationary point $K = K_c $,
\begin{equation}
e^{K_c }  = f\left( {e^{K_c } } \right) ,
\label{eq:17}
\end{equation}
corresponding to the transition point. At $K > K_c$ $K_l  \to \infty $, and at $K < K_c $ $K_l  \to 0$. When $\left| {K - K_c } \right| \ll K_c $, then for sufficiently small $l$ $K_l $ varies slowly, 
\begin{eqnarray}
e^{K_l  - K_c}  \approx 1 + \kappa ^l \left( K - K_c \right),\label{eq:18}
\\
\kappa  \equiv f'\left( e^{K_c }\right) = \nonumber
\\
B\frac{\left( e^{K_c}  - e^{K_c\left(n - 1\right)/n}\right)\left(e^{K_c /n}  + q - 1\right)}{\left(e^{K_c }  - 1\right)\left(e^{K_c }  + q - 1 \right)} < B, \label{eq:19}
\end{eqnarray}
until it becomes much larger or much less than $K_c $. The condition for validity of Eq. (\ref{eq:18}) is
\begin{equation}
l < l_c  \equiv \ln \frac{const}{\left|K - K_c\right|}/\ln \kappa 
\label{eq:20}
\end{equation}
Here the constant is determined from Eq. (\ref{eq:18}) by the condition $K_{l_c } \sim K_c $ at $K < K_c$ and by the condition $K_{l_c}  \gg K_c $ at $K > K_c $.
At $l > l_c $ it follows from Eq. (\ref{eq:14})
\begin{eqnarray}
 K < K_c,
 {\rm   }
  K_l  \approx qn^{- 1/\left( m - 1 \right)} \left( K_{l_c} n^{1/\left(m - 1 \right) /q} \right)^{m^{l - l_c}} \label{eq:21} \\ 
 K > K_c,
 {\rm   }
 \exp K_l  \approx m^{n/ \left( n - 1 \right)} \left( m^{- n/\left(n - 1\right)} \exp K_{l_c } \right)^{n^{l - l_c }} \nonumber
\end{eqnarray}
Using Eqs. (\ref{eq:18})-(\ref{eq:21}) one can show that near transition the potential (\ref{eq:16}) has singular part proportional to $B^{ - l_c } \sim \left| K - K_c \right|^{2 - \alpha }$ with the heat capacity index
 \begin{equation}
\alpha  = 2 - \frac{\ln B}{\ln \kappa }
\label{eq:22}
\end{equation}
Thus at $K > K_c $ and $\left| K - K_c \right|\sim K_c $ assuming in Eq.  (\ref{eq:16})
\[
g_k  \approx \left\{ \begin{array}{l}
 g_c  \equiv \left( e^{K_c }  - 1 \right)^B /\left( e^{K_c /n}  - 1 \right)^n,{\rm   }l < l_c  \\ 
 g_\infty   \equiv \left( q^{m - 1 /n} \right)^n,{\rm   }l > l_c  \\ 
 \end{array} \right.
\]
we get
\[
F \approx  - \frac{\bar z}{2\left( B - 1 \right)}T\left[ g_c  + B^{ - l_c} \left( g_\infty   - g_c \right) \right] .
\]

At $h = 0$ it is also easy to find the correlation function for the basic spins $\sigma ,{\rm  }\sigma '$
\[
G = \left\langle \delta _{\sigma ,1} \delta _{\sigma ',1}  \right\rangle  - \left\langle \delta _{\sigma ,1}  \right\rangle \left\langle \delta _{\sigma ',1} \right\rangle .
\]
On the $l$-th level lattice we have
\[
G_l=\frac{Z_l \left( {1,1} \right)}{Z_l^f } - \left( \frac{\sum\limits_\sigma  {Z_l \left( \sigma ,1 \right)}}{Z_l^f } \right)^2  = \frac{q - 1}{q^2 } \frac{e^{K_l }  - 1}{e^{K_l }  + q - 1}
\]
At $K < K_c$, $l > l_c $ it follows from Eq.  (\ref{eq:21})
\begin{eqnarray*}
G_l\approx K_l /q^3  \approx q^{-2} n^{- 1/\left( m - 1 \right) } \left( K_{l_c } n^{1/\left( m - 1 \right)}/q \right)^{m^{l - l_c }} \\
\sim \exp \left(- m^l /\xi \right)
\end{eqnarray*}
where the correlation length
\begin{equation}
\xi \sim m^{l_c } \sim \left( K_c  - K \right)^{ - \nu },
\qquad
\nu  = \frac{\ln m}{\ln \kappa }
\label{eq:23}
\end{equation}
The correlation length index $\nu $ obeys the scaling relation
\[
d\nu  = 2 - \alpha 
\]
At $K > K_c $ from the second relation in Eq. (\ref{eq:21}) it follows that $G_l $ goes to a constant at large $l$,
\[
G_l  \approx \left( q - 1 \right)/q^2  + A\exp \left(  - const \cdot n^{l - l_c }  \right) ,
\]
but in this case the characteristic length of $G_l $ variations is also proportional to $m^{l_c } $, as $n^{l - l_c }  = \left( m^l /m^{l_c} \right)^{d - 1} $.
Thus on the lattices considered the Potts model has always power low singularities of the heat capacity and the correlation length, with exponents obeying the usual scaling relation.  In the next section we consider the critical behavior of the order parameter and the critical susceptibility.

\section{Order parameter and critical susceptibility}
The spontaneous order parameter of the Potts model is
\begin{equation}
\mu  =  \lim \limits_{l \to \infty } \frac{ qN_l^{ - 1} 
\sum\limits_{i = 1}^{N_l  - 1} { \left\langle \delta _{\sigma _i ,1} \right\rangle}  - 1} { q - 1 }
\label{eq:24}
\end{equation}
Here  $\left\langle {\delta _{\sigma ,1} } \right\rangle $ means the average with boundary condition $\sigma  = 1$ on the basic sites, that is
\begin{equation}
\sum\limits_{i = 1}^{N_l  - 1} {\left\langle \delta _{\sigma _i ,1}  \right\rangle }  = \left. \frac{1}{Z_l^{\left( 1 \right)} }\frac{\partial Z_l^{\left( 1 \right)} }{\partial h} \right|_{h = 0}  \equiv \frac{\dot Z_l^{\left( 1 \right)} }{Z_l^{\left( 1 \right)} }
\label{eq:25}
\end{equation}
Just the use of such symmetry breaking boundary conditions allows to obtain a nonzero $\mu $ at $K > K_c $ and $h = 0$. 
Indeed in zero field and for free and periodic boundary conditions $\left\langle \delta _{\sigma _i, 1}  \right\rangle  = 1/q$ due to the permutation symmetry of $\sigma $ values. This results, particularly, in the relations 
\[
\dot Z_l^{\left( f \right)}  = N_l Z_l^{\left( f \right)} /q
\qquad
\dot Z_l^{\left( p \right)}  = \left( {N_l  - 1} \right)Z_l^{\left( p \right)} /q{\rm  }
\]
Inserting here Eqs (\ref{eq:11}), (\ref{eq:12}), we get
\begin{eqnarray}
\dot \lambda _{ + l}  = \left( {N_l  - 1} \right)\lambda _{ + l} /q
\label{eq:26}\\
\dot \lambda _{ - l}  + \left( q - 2 \right)\dot c_l  = \left( N_l  - 1 \right)\left(q - 1 \right)\lambda _{ - l} /q
\nonumber
\end{eqnarray}
Here $\dot \lambda _{ \pm l} $ are the field derivatives of the eigenvalues in Eqs. (\ref{eq:7}) at $h = 0$
\[
\dot \lambda _{ \pm l}  = \frac{1}{2}\left( \dot a_{1l}  + \dot a_{2l}  \right) \pm \frac{1}{2q}\left[ {\left( 2 - q \right)\left( \dot a_{1l}  - \dot a_{2l}  \right) + 2\left( q - 1 \right)\dot b_l } \right] .
\]
From Eqs. (\ref{eq:13}) (\ref{eq:24}) - (\ref{eq:26}) we have
\begin{eqnarray}
\mu  = \mathop {\lim }\limits_{l \to \infty } \frac{{qN_l^{ - 1} \varphi _{ + l}  + q - 2}}{{2\left( {q - 1} \right)}} \label{eq:27}
\\
\varphi _{ + l}  = \left[ {\dot a_{1l}  - \dot a_{2l}  - \left( {q - 2} \right)\dot c_l } \right]/a_{1l}. 
\nonumber
\end{eqnarray}
Let us introduce one more linear combination of the derivatives linearly independent with $\varphi _{ + l} $ and those in Eqs. (\ref{eq:26}),
\[
\varphi _{ - l}  = \left( \dot a_{1l}  - \dot a_{2l}  + q\dot c_l \right)/a_{1l} \]
Then, differentiating Eq. (\ref{eq:9}) over $h$ and using Eq. (\ref{eq:26}) we get the reccurence relations for the vector $ \bm \varphi _l  = \left( {\varphi _{ + l} ,\varphi _{ - l} } \right)$
\begin{equation}
\bm \varphi _l  = \hat T_l \bm \varphi _{l - 1}  + \bm u_l ,
\label{eq:28}
\end{equation}
where
\begin{eqnarray}
u_{ + l}  = 1 - n + \frac{q - 2}{q^2 }n\left( N_{l - 1}  - 1 \right)
\nonumber
\\
\times \left[ \left( q - 2 \right)\left( m - e_l \right) - 2\left( q - 1 \right)\left(m - 1 \right)e^{ - K_l /n} \right] 
 \label{eq:29}
\\
u_{ - l}= \frac{e'_l}{e_l}u_{ + l}+ \left( N_l  - 1 \right)\left( 1 - \frac{e'_l }{e_l} \right)  \label{eq:30}
\\
\hat T_l  = \frac{n}{q} \cdot \left( \begin{array}{*{20}c}
{e_l \left[ 2 + \left( q - 2 \right)m\vartheta _l \right]} & {e_l \left( q - 2 \right)\left( 1 - m\vartheta _l \right)}  \\
{2e'_l \left( 1 - m\vartheta _l \right)} & {e'_l \left( q - 2 + 2m\vartheta _l \right)} \end{array}\right) \label{eq:31}
\\
e_l  \equiv \exp \left( K_{l - 1}  - K_l \right),
\qquad
e'_l  \equiv \exp \left( K_{l - 1}  - K_l \right), \nonumber\\
\vartheta _l  \equiv \left[ \exp \left( K_l /n \right) - 1 \right]/\left[ \exp \left( K_{l - 1} \right) - 1 \right].
\label{eq:32}
\end{eqnarray}
The solution of Eq. (\ref{eq:27}) is
\begin{equation}
\bm \varphi _l  = \hat T_l \hat T_{l - 1} ...\hat T_1 \bm \varphi _0  + \sum\limits_{k = 1}^{l - 1} {\hat T_l \hat T_{l - 1} ...\hat T_{k + 1} \bm u_k }  + \bm u_l 
\label{eq:33}
\end{equation}
where $\bm \varphi _0  = \left( 1,1 \right)$.

Let us consider the asymptotics of $\bm \varphi _l $ at $l \to \infty $ near the transition. Here $\hat T_l $ can be approximated as
\begin{equation}
\hat T_l  \approx \left\{ \begin{array}{l}
\hat T_c  \equiv \mathop {\lim }\limits_{K_l  \to K_c } \hat T_l ,{\rm    }l < l_c  \\ 
 \hat T_\infty   \equiv \mathop {\lim }\limits_{l \to \infty } \hat T_l ,{\rm    }l > l_c  \\ 
 \end{array} \right.
\label{eq:34}
\end{equation}
Then	
\begin{eqnarray*}
\bm \varphi _l  \approx \hat T_\infty ^{l - l_c } \hat T_c^{l_c } \left[ \bm \varphi _0  + \sum\limits_{k = 1}^{l_c } {\left( {B\hat T_c^{ - 1} } \right)^k } \bm u_c   \right] \\ 
 + \sum\limits_{k = l_c  + 1}^l {B^k \hat T_\infty ^{l - k}} \bm u_\infty,  
 \end{eqnarray*} 
 or in simpler form 
 \begin{eqnarray}
 \bm \varphi _l  \approx \hat T_\infty ^{l - l_c } \hat T_c^{l_c } \left( \bm \varphi _0  - \bm \varphi _c  \right) + B^{l_c } \hat T_\infty ^{l - l_c } \bm \varphi _c \nonumber\\
 + B^l \left[ \hat I - \left( B^{ - 1} \hat T_\infty  \right)^{l - l_c } \right] \bm \varphi _\infty   \label{eq:35}
 \end{eqnarray}
Here
\begin{eqnarray*}
u_c  =  \lim \limits_{l \to \infty } \lim \limits_{K_l  \to K_c } \bm u_l B^{ - l} ,
\qquad
\bm u_\infty   = \mathop {\lim }\limits_{l \to \infty } \bm u_l B^{ - l} 
\\
\bm \varphi _c  \equiv \left( \hat I - B^{ - 1} \hat T_c  \right)^{ - 1} \bm u_c \qquad
\varphi _\infty   \equiv \left( {\hat I - B^{ - 1} \hat T_\infty  } \right)^{ - 1} \bm u_\infty  
\end{eqnarray*}

From Eqs. (\ref{eq:29})-(\ref{eq:32}) we get
\begin{eqnarray}
\hat T_c  = \frac{n}{q} \cdot \left( {\begin{array}{*{20}c}
{e_c \left[ 2 + \left( q - 2 \right)m\vartheta _c  \right]} & {e_c \left( q - 2 \right)\left( 1 - m\vartheta _c  \right)}  \\
{2\left( 1 - m\vartheta _c  \right)} & {\left( q - 2 + 2m\vartheta _c  \right)}  \\
\end{array}} \right) \label{eq:36}
\\
e_c  = \exp \left[ {K_c \left( {n - 1} \right)/n} \right],\nonumber
\\
\vartheta _c  = \left[ \exp \left( K_c /n \right) - 1 \right]/\left[ \exp \left( K_c \right) - 1 \right]. \nonumber
\\
\varphi _{c + }  =  - \frac{2}{\bar z}\frac{q - 2}{q},
\qquad
\varphi _{c - }  =  - \frac{2}{{\bar z}}\left( {1 - 2e^{ - K_c } \frac{{q - 1}}{q}} \right).
\label{eq:37}
\end{eqnarray}

It is important for the following that the eigenvalues of the matrix $\hat T_c $ in Eq. (\ref{eq:36}) are real and are less than $B$ all $m \ge 2$, $n \ge 2$, $q > 0$. Indeed, the largest eigenvalue of $\hat T_c $ is
\begin{eqnarray}
2\lambda _{\max } /n = m +\varepsilon -\rho +\sqrt{\left( m +\varepsilon  - \rho \right)^2  - 4m\varepsilon} 
\label{eq:38}
\\
\rho  \equiv \frac{2}{q}\left( 1-\varepsilon \right)\left[ me^{K_c /n}  - e^{K_c }  + \frac{q - 2}{2}\left( m - 1\right)\right]\nonumber
\\
\varepsilon\equiv e_c \vartheta _c  < 1 \nonumber
\end{eqnarray}
Using Eq. (\ref{eq:17}) for $K_c $ one can show that $0 < \rho  < \left( {m - 1} \right)\left( 1-\varepsilon \right)$ at all $m \ge 2$, $n \ge 2$, $q > 0$ and this guarantees the reality of $\hat T_c $ eigenvalues and the inequality $\lambda _{\max } < B$.

The expressions for $\hat T_\infty$  and $u_\infty  $ differ in ordered and disordered phases. Thus at $K < K_c $ it follows from Eqs.  (\ref{eq:21}), (\ref{eq:31}), (\ref{eq:32}), (\ref{eq:34}) 
\[
\hat T_\infty   = \frac{n}{q} \cdot \left( \begin{array}{*{20}c}
   2 & {q - 2}  \\
   2 & {q - 2}  \\
\end{array} \right),
\qquad
u_\infty   =  - \frac{q - 2}{q} \cdot \frac{m - 1}{m} \cdot \frac{2}{\bar z} \cdot \left( \begin{array}{l}
1 \\ 
1 \\ 
 \end{array} \right) .
\]

At $K > K_c $
\[
\hat T_\infty   = \left( {\begin{array}{*{20}c}
B & 0  \\
0 & 0  \\
\end{array}} \right),
\qquad
\bm u_\infty   = \left( \begin{array}{l}
 0 \\ 
 1 \\ 
 \end{array} \right) ,
\]
so the part in Eq. (\ref{eq:35}) containing $\bm u_\infty  {\rm  }\left(\bm \varphi _\infty \right)$ is zero and we have
\begin{eqnarray*}
\varphi _{ + l}  = B^{l - l_c } \bm e_ +  \hat T_c^{l_c } \left( \bm \varphi _0  - \bm \varphi _c  \right) + B^l \varphi _{c + }  
\\
\approx N_l \left[ const\left( \lambda _{\max } /B \right)^{l_c }  - \left( q - 2 \right)/q \right] .
\end{eqnarray*}

Thus we have for the order parameter in Eq. (\ref{eq:27})
\begin{equation}
\mu \sim \left( \lambda _{\max } /B \right)^{l_c } \sim \left( {K - K_c } \right)^\beta , 
\qquad
\beta  = \frac{\ln \left( B/\lambda _{\max }  \right)}{\ln \kappa } > 0
\label{eq:39}
\end{equation}

Let us now consider the critical susceptibility at $h = 0$
\begin{eqnarray}
\chi  \equiv \lim \limits_{l \to \infty } N_l^{ - 1} \left[ \frac{\ddot Z_l^{\left( f \right)} }{Z_l^{\left( f \right)} } - \left( \frac{\dot Z_l^{\left( f \right)}}{Z_l^{\left( f \right)}}\right)^2  \right] \nonumber
\\
=  \lim \limits_{l \to \infty } \left( {N_l^{ - 1} \psi _{ + l}  - N_l /q^2 } \right) + 2/q
\label{eq:40}
\end{eqnarray}
Here we used the relations in Eq. (\ref{eq:26}) and introduced
\[
\psi _{ + l}  = \left[ \ddot a_{1l}  + \left( q - 1 \right)\left( \ddot a_{2l}  + 2\ddot b_l  \right) \right]/q\lambda _{ + l} .
\]
Defining
\[
\psi _{ - l}  = \left[ \left( q - 1 \right)\left( \ddot a_{1l}  - 2\ddot b_l  \right) + \ddot a_{2l}  + q\left( q - 2 \right)\ddot c_l \right]/q\lambda _{ - l} 
\]
and differentiating Eq. (\ref{eq:9}) twice over $h$ we get the recurrence relations for the vector
$\bm \psi _l  = \left( \psi _{ + l} ,\psi _{ - l}  \right)$
\begin{equation}
\bm \psi _l  = \hat P_l \bm \psi _{l - 1}  + \bm v_l 
\label{eq:41}
\end{equation}
where
\begin{eqnarray}
\hat P_l  = B \cdot \left( \begin{array}{*{20}c}
{x_l } & {\left( 1 - x_l  \right)/\left( q - 1 \right)}  \\
{\left( q - 1 \right)y_l } & {1 - y_l }  \\
\end{array} \right) , \label{eq:42}
\\
x_l  = \left( e^{K_l /n  + q - 1} \right)\frac{e^{K_l \left( n - 1 \right)/n}  + q - 1}{e^{K_l}   + q - 1}, \nonumber
\\
y_l  = \left( e^{K_l /n}  + q - 1 \right)\frac{e^{K_l \left( n - 1 \right)/n}  - 1}{e^{K_l }  - 1}, \nonumber
\end{eqnarray}
and $\bm v_l $ at $\left| K - K_c  \right| \sim K_c $ and large  $l$ has the form
\begin{equation}
\bm v_l  = \frac{B - 1}{B}\frac{N_l^2 }{q^2 }\left( 
\begin{array}{l}
{\rm   }1 \\ 
q - 1 \\ 
\end{array} \right) + {\bm c}\lambda _{\max }^{2l}  + O\left( N_l  \right) .
\label{eq:43}
\end{equation}
The matrices $\hat P_l $ has eigenvalues $B$ and $B\left( {x_l  - y_l } \right) < B$. The contribution to $\bm v_l $ proportional to $N_l^2 $ is the right eigenvector for all $\hat P_l $ corresponding to the eigenvalue $B$. Using the last circumstance and the approximation for $\hat P_l $, analogous to that in Eq. (\ref{eq:34}), we get from Eqs. (\ref{eq:40}) - Eq. (\ref{eq:43}) near transition
\begin{eqnarray}
\chi \sim \left( \lambda _{\max }^2 /B \right)^{l_c} \sim \left| K - K_c  \right|^{ - \gamma },\nonumber
\\
\gamma  = \left( 2\ln \lambda _{\max }  - \ln B \right)/\ln \kappa .
\label{eq:44}
\end{eqnarray}
Apparently the usual scaling relation is valid,
\[
\alpha  + 2\beta  + \gamma  = 2.
\]
Thus the Potts model on the lattices considered always undergoes a second-order phase transition with power-law singularities of thermodynamic parameters.

\section{Critical indexes}
Here we consider the dependence of critical indexes on the lattice parameters and number of Potts states $q$. The inequalities $\kappa  < B$, $\lambda _{{\rm max}}  < B$ (see Eqs. (\ref {eq:19}), (\ref {eq:38})) and scaling relations results in the following boundaries for the values of critical indexes in Eqs. (\ref {eq:22}), (\ref {eq:23}), (\ref {eq:39}), (\ref {eq:44})
\[
\nu  > 1/d,
\qquad
\alpha  < 1,
\qquad
\beta  > 0,
\qquad
\gamma  > 1.
\]
When $q \to \infty $ the indexes go to the limiting values in these inequalities. Indeed for $q \gg 1$ it follows from Eqs. (\ref {eq:17}), (\ref {eq:19}), (\ref {eq:38}) $K_c  \approx \left( 2/\bar z\right)\ln q$, $\kappa  \approx B$, $\lambda _{ max}  \approx B\varepsilon $, so 
\[
\nu  \approx 1/d,
\qquad
\alpha  \approx 1,
\qquad
\beta  \approx q^{-2/\bar zn} \ln B
\qquad
\gamma  \approx 1.
\]
Simple expressions for the indexes can be obtained for $n \to \infty $ $\left( {d \to \infty } \right)$, when $K_c  \approx qn^{ - 1/\left( m - 1 \right)}$,  $\kappa  \approx m$, $\lambda _{max}  \approx n$, so
\[
\nu  \approx 1,
\qquad
\alpha  \approx 2 - d,
\qquad
\beta  \approx 1,
\qquad
\gamma  \approx d - 2.
\]
Also at $m \to \infty$ $\left( d \to 1 \right)$, when $K_c  \approx \frac{n}{n - 1}\ln m$, $\kappa  \approx n$, $\rho  \approx qm^{\frac{n - 3}{n - 1}} /6$ ,  $\lambda _{\max}  \approx B\left(1 - \rho /m \right)$, we get
\begin{eqnarray*}
\nu  \approx \frac{1}{d - 1},
\qquad
\alpha  \approx \frac{d - 2}{d - 1},
\qquad
\gamma  \approx \frac{d}{d - 1},
\\
\beta  \approx \frac{q}{6m^{2/\left( n - 1 \right)} \ln n}.
\end{eqnarray*}

\begin{figure*}
\includegraphics{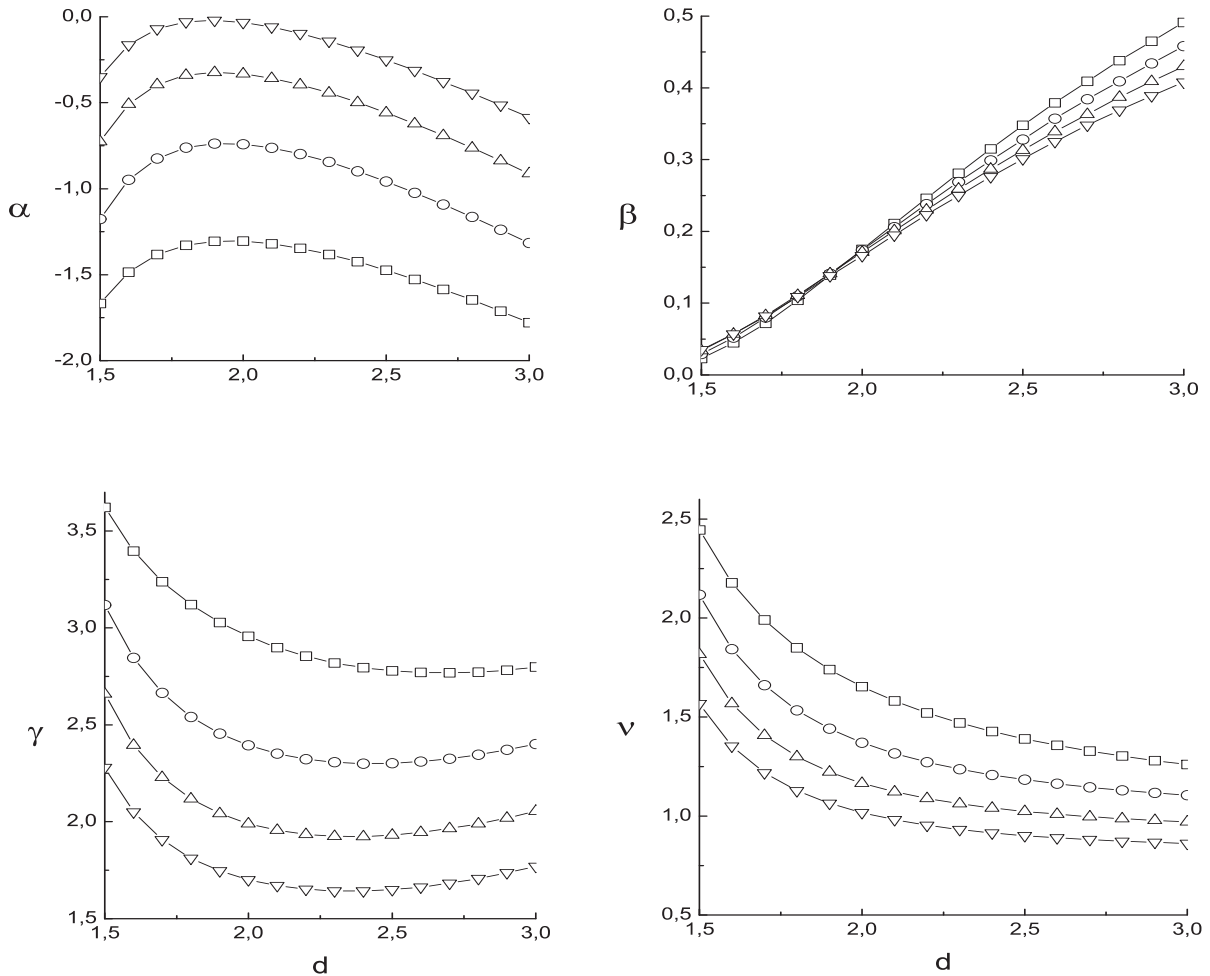}
\caption{\label{Fig.2} The dependence of the critical indexes of the $q$-state Potts model on the hierarchical lattices with $\bar z = 2.5$ on the fractal dimension $d$; $\square-q=1$, $\bigcirc-q=2$, $\triangle-q=4$, $\nabla-q=8$.}
\end{figure*}

Note that in these limiting cases only $\beta $ depends on $q$ and only when it is small. In general, considering the indexes as the functions of the observable fractal parameters $d$, $\bar z$, one can find that their values only slightly depend on the average coordination number in the interval $2.5 < \bar z < 4$. Fig. (\ref{Fig.2}) presents the indexes' dependencies on fractal dimension in the physically realizable interval $1.5 <d < 3$ at $\bar z =2.5$ and various $q$.
At $q=1$ the indexes describe the critical anomalies for the percolation of bonds randomly scattered over the hierarchical lattice with probability $p=1-\exp^{-K}$, cf. Ref. \onlinecite{31}. The equation for the percolation threshold $p_c=1-\exp^{-K_c}$ follows from Eq. (\ref{eq:17})
\[
p_c  = 1 - \left( {1 - p_c^m } \right)^n .
\]

Let us note that all variety of the indexes of Potts model on fractals, illustrated in Fig. (\ref{Fig.2}), has some definite features. Thus $\alpha < 0$ for not very large $q$, the susceptibility index is anomalously large ($\gamma \ge 1.7$), $\nu$ shows the monotonous diminishing while $\beta$ grows monotonously with a growth of the fractal dimension $d$. Note also a monotonous dependence of the indexes on $q$, the only exception is $\beta$. Probably, these properties of the indexes would be preserved in other spin models on the hierarchical lattices at small enough $d$.

We should also note that the indexes $\alpha $ and $\beta$ at $m=2$ and some $n$ and $q$ were determined earlier in Ref. \onlinecite{16}. The expressions for $\alpha $ and $\beta$ obtained here (Eqs. (\ref{eq:22}) and (\ref{eq:39})) give the same values up to the calculation errors.

\section{On a criterion of transition order transformation}

Considering hierarchical lattices as subclass of a wider set of inhomogeneous fractals which includes such real objects as porous materials and percolation clusters in dilute crystals, one may attempt to draw some conclusions about their essential structural characteristics, which bring out the transformation of the first-order transitions into the second-order ones. Thus one can note that on the hierarchical fractals considered here this phenomenon does not depend on fractal dimension, at all $1<d<\infty$ the second-order transition appears. One can suppose that the feature that accounts for the transition order transformation in these lattices is their small average coordination number $\bar z <4$. Analogous situation appears in dilute $2d$ quadratic lattice, where dilution makes $\bar z <4$ leaving the fractal dimension of the largest cluster $d=2$ up to the percolation threshold  \cite{6}. So, quite similarly, an arbitrary small concentration of vacancies transforms a first-order transition into a second-order one in this case, cf. Refs. \onlinecite{10, 11, 33}. At the same time, in the model of random media with $\bar z>4$ a first-order transition can be smeared but does not become a second-order one \cite{25}.

It seems rather probable that the condition $\bar z <4$ could be a criterion of the transition order transformation in a rather large class of inhomogeneous systems with short-range interactions. Particularly, in dilute models on the simple cubic lattice ($z=6$) with vacancies' concentration $1-x$ one may roughly estimate the average coordination number in the percolation cluster as $\bar z =6x$. Then from the condition $\bar z <4$ it follows that a second-order transition would appear at $x<2/3$. This estimate are close to the result of numerical studies, $x<0.7\pm 0.05$ \cite{12, 13}.

Qualitatively the existence of the threshold coordination number for the transition order transformation could be explained by the absence of the macroscopic regions with the largest coordination number in the systems with low connectivity (low $\bar z$). This makes energetically unfavorable the appearance of the macroscopic regions of the ordered phase near the transition points. In this case instead of creation of ordered regions only spreading of the order parameter correlations is possible, which is characteristic for the second-order transitions and causes the singularities of the thermodynamic parameters. Quite contrarily, in the structures with large $\bar z$ the phenomenological scenario of Ref. \onlinecite{17} could be realized, which supposes the appearance and growth in volumes and numbers of macroscopic ordered regions near transition. Then both in the  ordered regions and in the disordered ones the order parameter correlation length stays finite and the second-order singularities are absent. Yet in this case the first-order jumps can completely vanish at large enough disorder \cite{25}.

\section{Discussion and conclusions}

The results obtained here can be used to describe (more or less quantitatively) the critical anomalies near phase transitions described by the Potts model in the percolation clusters of dilute crystals or in substances confined in porous media, which have fractal structures similar to those of the hierarchical lattices considered. This possibility seems rather reasonable for $q \le 10$ as then $\alpha < 0$ in conformity with the exact condition for this index in random systems \cite{32}.
The analytical expressions for critical indexes obtained here make their comparison with experimental data quite operational.

We may note that the specific feature of the critical indexes obtained here, which are determined mainly by the fractal dimension and only weakly depend on $\bar z$ at $2.5 < \bar z <4$, allows to explain their slight variations with the change of defect concentration in dilute crystals \cite{9,10,11,12,13}. Indeed, the fractal dimension of percolation cluster, where transition takes place in such crystals, is almost constant up to the percolation threshold \cite{6}, so  slight changes of critical indexes result only from their weak dependence on the average coordination number. At the same time, one can expect more essential variations of the critical indexes when the substance undergoing transition is confined in porous matrices as their fractal dimension can vary in a wide interval \cite{6}.

Unfortunately now the comparison of the indexes obtained with their experimental values is impossible due to the absence of the detailed data on the critical anomalies near the inhomogeneity-induced second-order transitions. Mainly this is the consequence of the fact that only the recent theoretical works \cite{10,11,12,13,14,15} have established the predetermination of this phenomenon. Meanwhile, of old the numerous experiments are known, which find the second-order transitions in crystals which should have the first-order ones according to the Landau theory of phase transitions \cite{24}. In the light of the Refs. \onlinecite{10,11,12,13,14,15} results one can consider such experimental data not as curiosity, but as a result of the presence in crystals of a considerable amount of defects and impurities. The characteristic examples are  $Nb_3Sn$ and $V_3Si$ crystals \cite{34} in which the distinction of the transitions from the second-order ones were established only on the samples of sufficiently high quality \cite{35}.

In conclusion we may note that, very probably, the inhomogeneity-induced second-order transitions are not the specific feature of the Potts model studied in Refs. \cite{10,11,12,13,14,15,16} and in the present work. It is naturally to suppose that this phenomenon can exist at all transitions to a subgroup of high-symmetry phase, which in ideal crystals are described by the Landau potential with a cubic invariant of order parameter components \cite{24}.
At the same time, there are neither experimental nor theoretical evidences of the appearance of this phenomenon when the relation group-subgroup between the symmetries of disordered and ordered phases is absent, that is for so-called reconstructive first-order transitions \cite{24}. Probably, in this case the appearance of the inhomogeneity-induced second-order singularities is impossible.

\begin{acknowledgments}
This work was made under support from INTAS, grant 2001-0826, and RFBR, grant 04-02-16228.
I gratefully acknowledge useful discussions with V. P. Sakhnenko, V. I. Torgashev, V. B. Shirokov, M. P. Ivliev and E. D. Gutlyanskii.
\end{acknowledgments}

\end{document}